\documentstyle[sprocl,psfig,epsfig]{article}
\def\be{\begin{equation}}
\def\ee{\end{equation}}
\def\bea{\begin{eqnarray}}
\def\eea{\end{eqnarray}}
\def\simlt{\stackrel{<}{{}_\sim}}
\def\simgt{\stackrel{>}{{}_\sim}}

\begin{document}
\begin{flushright}
IFT/97-18\\
hep--ph/9711470 \\
\end{flushright}
\vskip 0.8cm
\title{CONSTRAINTS ON LOW ENERGY SUPERSYMMETRY\footnote{To appear
in the Proceedings of the International Workshop on 
Quantum Effects in the MSSM, Barcelona, September 1997.}}
\author{PIOTR H. CHANKOWSKI}
\address{Institute of Theoretical Physics, Warsaw University, 
Ho\.za 69, 00-681 Warsaw, Poland.}
\maketitle\abstracts{We review the available constraints on the low energy 
supersymmetry. The bulk of the electroweak data is well screened from 
supersymmetric loop effects, due to the structure of the theory, even with 
superpartners generically light, ${\cal O}(M_Z)$. The only exception are the 
left-handed squarks of the third generation which have to be 
$\simgt {\cal O}(300$ GeV) to maintain the success of the SM in describing 
the precision data. The other superpartners can still be light, at their 
present experimental mass limits. As an application of the derived constraints
(supplemented by the requirement of ``naturalness'') we discuss the 
predictions for the mass of the lighter MSSM Higgs boson in specific
scanarios of supersymmetry breaking.}

\section{Introduction}
Supersymmetry offers an interesting solution to the well known hierarchy 
puzzle of the SM and, moreover, has several other theoretical and 
phenomenological (gauge coupling unification) virtues. 
The supersymmetry breaking scale (often it can be defined only in some 
average sense) i.e. the scale of the mass spectrum of the superpartners 
provides the necessary cut-off to the SM. Two immediate 
and most important remarks about the superpartner spectrum are the following 
ones: if supersymmetry is to cure the hierarchy problem that scale is 
expected to be not much above the electroweak scale. On the other hand, it 
is totally unknown in detail, as we do not have at present any realistic model 
of supersymmetry breaking. Therefore, the Minimal Supersymmetric Standard Model
(MSSM) is a very well defined theoretical framework but contains many free 
parameters: superpartner soft masses and their dimensionful couplings. 

It is, therefore, very interesting to discuss the question to what extent the
superpartner spectrum can manifest itself through virtual (loop) effects 
on the electroweak observables. Do very high precision measurements of the 
electroweak observables  provide  us with a tool to see supersymmetric 
effects indirectly or, at least, to put stronger limits on its spectrum? 
We remember the important r\^ole played by precision measurements in seeing,
indirectly, some evidence for the top quark long ago its direct discovery and
with the mass quite close to its measured mass. Also, the present level of
precision makes the electroweak measurements to some extent sensitive even 
to the Higgs boson mass, although the dependence is only logarithmic. With
supersymmetric corrections the situation is different. The dependence on the 
top quark (and Higgs) mass in the SM is due to {\sl nondecoupling} of heavy 
particles which get their masses through the mechanism of spontaneous symmetry
breaking. The soft SUSY breaking is explicit and the Appelquist-Carazzone 
theorem \cite{APCA} applies to the superpartner spectrum. Thus, 
supersymmetric virtual effects dissapear at least as ${\cal O}(1/M_{SUSY})$. 
Nevertheless, several interesting questions can be discussed.

\section{Constraints on general MSSM}

The bulk of the electroweak precision measurements ($M_W$, $Z^0$-pole 
observables, $\nu e$, $ep$ scattering data, etc.) shows that the global 
comparison of the SM predictions with the data is impressive (for more details 
see e.g.\cite{HOLL}). The simplest interpretation of the success of the SM 
within the MSSM is that the superpartners are heavy enough to decouple from 
the electroweak observables. Explicit calculations show that this happens if 
the common supersymmetry breaking scale is $\geq {\cal O}(300-400)$ GeV 
\cite{ABC,KANE2}. This is very important as such a scale of supersymmetry 
breaking 
is still low enough for supersymmetry to cure the hierarchy problem. However, 
in this case the only supersymmetric signature at the electroweak scale and 
just above it is the Higgs sector with a light, $M_h\leq {\cal O}(150)$ GeV, 
Higgs boson. This prediction is consistent with global fits in the SM which 
give $M_h\approx130^{+130}_{-70}$ GeV (the 95$\%$ C.L. upper bound is around 
470 GeV) \cite{ELLIS,LEPEWWG97}. We can, therefore, conclude at this point 
that the supersymmetric extension of the SM, with all superpartners 
$\geq {\cal O}(300)$ GeV, is phenomenologically as succesful as the SM itself 
and has the virtue of solving the hierarchy problem. Discovery of a light 
Higgs boson is the crucial test for such an extension.

The relatively heavy superpartners discussed in the previous paragraph are 
sufficient for explaining the success of the SM. But is there a room for some 
light superpartners with masses ${\cal O}(M_Z)$ or even below? This question 
is of great importance for LEP2. Indeed, a closer look at the electroweak 
observables shows that the answer to this question is positive. The dominant 
quantum corrections to the electroweak observables are the so-called "oblique"
corrections to the gauge boson self-energies. They are economically summarized 
in terms of the $S,T,U$ \cite{STU} parameters 
\begin{eqnarray}
\alpha S\sim \Pi^\prime_{3Y}(0)=\Pi^\prime_{L3,R3} +\Pi^\prime_{L3,B-L}
\label{eqn:Spar}
\end{eqnarray}
(the last decomposition is labelled by the
$SU_L(2)\times SU_R(2)\times U_{B-L}(1)$ quantum numbers \cite{JAPP})
\begin{eqnarray}
\alpha T\equiv\Delta\rho\sim\Pi_{11}(0)-\Pi_{33}(0)
\end{eqnarray}
\begin{eqnarray}
\alpha U\sim \Pi^\prime_{11}(0)-\Pi^\prime_{33}(0)
\end{eqnarray}
where $\Pi_{ij}(0)$ $(\Pi^\prime_{ij}(0))$ are the (i,j) gauge boson 
self-energies at the zero momentum (their derivatives) and the self-energy 
contribution to the $S$ parameter originates from mixing between $W^3_\mu$ 
and $B_\mu$ gauge bosons. The parameters $S,T,U$ have important symmetry 
properties: $T$ and $U$ vanish when quantum corrections to the gauge boson 
self-energies leave unbroken "custodial" $SU_V(2)$ symmetry. The parameter 
$S$ vanishes if $SU_L(2)$ remains an exact symmetry \cite{JAPP}. 

In terms of the parameters $S$, $T$ and $U$ the ``new physics'' contribution
to the basic electroweak observables can be approximately written e.g. as 
\begin{eqnarray}
\delta M_W = {M_W\over2}{\alpha\over c_W^2-s_W^2}\left(
c^2_W T^{new} - {1\over2}S^{new}
+{c^2_W-s^2_W\over4s^2_W}U^{new}\right)
\label{eqn:stu_mw}
\end{eqnarray}
\begin{eqnarray}
\delta\sin^2\theta^{eff}_{lept} =-{s^2_Wc^2_W\over c^2_W -s^2_W}\left(
\alpha T^{new}-{\alpha\over4s^2_Wc^2_W}S^{new}\right)
\label{eqn:stu_sin}
\end{eqnarray}
where the parameters $M_W$, $c_W\equiv M_W/M_Z$, $s_W$, etc.
are computed in the 
SM (taking into account loop corrections) with some reference values of 
$m_t$ and $M_h$. $S^{new}$, $T^{new}$, $U^{new}$ contain only the 
contributions from physics beyond the SM. 

The success 
of the SM means that it has just the right amount of the $SU_V(2)$ breaking 
(and of the $SU_L(2)$ breaking), encoded mainly in the top quark-bottom quark 
mass splitting. Any extension of the SM, to be consistent with the precision 
data, should not introduce additional sources of large $SU_V(2)$ breaking.

\begin{figure}
\psfig{figure=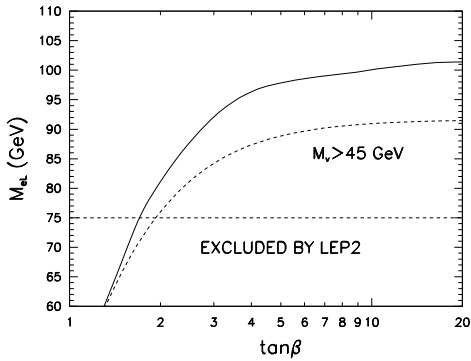,height=5.0cm}
\caption{Limits on degenerate charged left-handed sleptons from the fit
to the electroweak data (slid line)as a function of $\tan\beta$. Dashed 
line show the limit from the constraint $M_{\tilde\nu}>45$ GeV and the 
horizontal lines marks the present experimental limit. }
\label{fig:barc1}
\end{figure}

In the MSSM, there are potentially two new $SU_V(2)$ breaking effects
\cite{MYDR}. Firstly, masses of the  superpartners of the up and down-type 
left-handed fermions are splitted through the $D-$term contribution. 
\begin{eqnarray}
M^2_{\tilde f_{down}}-M^2_{\tilde f_{up}}\approx-\cos2\beta M^2_W
\label{eqn:leptsplit}
\end{eqnarray}
For example for light sleptons the splitting between the left-handed slepton 
and sneutrino masses becomes non-negligible. Assuming masses of all three
generation of sleptons to be the same and requiring $\Delta\chi^2<4$ for 
the MSSM fit to the electroweak data\cite{LEPEWWG97} one gets the lower bound
on the mass of the left-handed charged slepton shown in Fig. \ref{fig:barc1}.
This limit, for $\tan\beta\simgt2$ is still better than the limit from
the direct search (also marked on the plot) of sleptons at LEP2.

Twice as big (because of two generations of squarks instead of three 
species of sleptons but times the color factor 3)
effects of the $SU_V(2)$ violation are present also for 
light left-handed squarks from the first two generations\footnote{The direct 
Tevatron bounds, $M_{\tilde q}\simgt150$ GeV, for $\tilde q\neq\tilde t$, may 
not apply e.g. if $R-$parity is broken or if the gluino is heavier than 
${\cal O}(350)$ GeV.}. However in this case no limits similar to those for
sleptons can be obtained from the fit. The reason for that is a different
behaviour of the $S$ parameter (\ref{eqn:Spar}). Indeed, for the 
contribution of a doublet of left-handed sfermions, we find\cite{JAPP,MYDR}:
\begin{eqnarray}
S=\lambda{N_c\over12\pi}\log{M_{\tilde f_{down}}\over M_{\tilde f_{up}}}
\end{eqnarray}
where $N_c=1(3)$, $\lambda=-1(+1/3)$ for sleptons (squarks). Since for the
sfermions under consideration $M_{\tilde f_{down}}>M_{\tilde f_{up}}$ 
$S$ is negative for sleptons and therefore adds up to the positive contribution
of $T$ in eqs (\ref{eqn:stu_mw},\ref{eqn:stu_sin}). On the contrary, for
squarks $S$ is positive and compensates the effects of $T$. In principle, 
$T$ itself is closely related to an observable $\Delta\rho$ but the existing 
result\cite{VIL} $\rho=1.006\pm0.034$ is not accurate enough to constrain 
masses of the left-handed squarks (if heavier than 45 GeV).

Another, and far more important, source of $SU_V(2)$ violation is due
to the splitting between the left-handed stop and sbottom masses
\cite{HABER,MYDR}:
\begin{eqnarray}
M^2_{\tilde t_L}-M^2_{\tilde b_L}=m^2_t-m^2_b+\cos2\beta M^2_W
\label{eqn:tbspl}
\end{eqnarray}
This source of the $SU_V(2)$ violation cannot, however, be treated 
independently of the similar effects originating from the left-right mixing
of stops. Thus, in this case the fit depends on the physical masses 
$M^2_{\tilde t_1}$, $M^2_{\tilde t_2}$ of both stops and on their mixing
angle $\theta_{\tilde t}$:
\begin{eqnarray}
\tilde t_L = 
\cos\theta_{\tilde t}\tilde t_1-\sin\theta_{\tilde t}\tilde t_2, ~~~
\tilde t_R = \sin\theta_{\tilde t}\tilde t_1+\cos\theta_{\tilde t}\tilde t_2
\nonumber
\end{eqnarray}
Absolute (i.e. optimized with respect to 
$\theta_{\tilde t}$) lower limit derived from the fit to precision data on 
the mass of the heavier stop as a function of the mass of the lighter one is 
shown by solid lines for low and large values of $\tan\beta$.

\begin{figure}
\psfig{figure=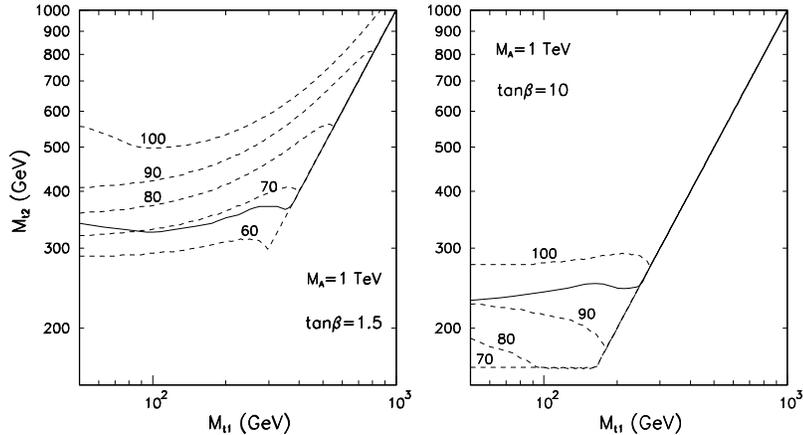,height=6.5cm}
\caption{Lower bounds on the heavier stop mass $M_{\tilde t_2}$, as a function
of $M_{\tilde t_1}$ for $\tan\beta=1.6$ and $\tan\beta=10$ from the fit
(solid lines) and from the assumed experimental limit on $M_h$
(dashed lines). A scan over the top quark mass and the 
top squarks mixing angle $\theta_{\tilde t}$ has been performed.
In this plot, $M_{\tilde t_1}<M_{\tilde t_2}$ by definition.}
\label{fig:barc2}
\end{figure}

Another interesting observation is that the upper limit on the mass of the 
lighter MSSM Higgs boson $h^0$ depends on the same set of parameters 
($M^2_{\tilde t_1}$, $M^2_{\tilde t_2}$, 
$\theta_{\tilde t}$)\cite{HIGGM}:
\begin{eqnarray}
M^2_{h^0}<M^2_{Z^0}\cos^22\beta+{3\alpha\over4\pi s^2_W}{m^4_t\over M^2_W}
\left[\log{M^2_{\tilde t_1}M^2_{\tilde t_2}\over m^4_t}
+ f\left(M^2_{\tilde t_1},M^2_{\tilde t_2},\theta_{\tilde t}\right)\right]
\label{eqn:mhha}
\end{eqnarray}
Therefore, similar absolute bound on $M^2_{\tilde t_2}$ as a function of 
$M^2_{\tilde t_1}$ can be derived from the limits on $M_h$ from direct
experimental searches. They are shown in Fig. \ref{fig:barc2} by dashed
lines for various assumed experimental limits on $M_h$. For low $\tan\beta$
where the MSSM lighter Higgs boson is SM-like the present experimental limit
$M_h^{SM}>77$ GeV can be used and gives the bound on $M^2_{\tilde t_2}$ which 
is already stronger than the one obtained from the  precision data. For
large $\tan\beta$ the limit from the fit is lover compared to the low 
$\tan\beta$ case (due to cancellation of $m_t^2$ and $M^2_W$ terms in eq. 
(\ref{eqn:tbspl})) but, nevertheless, still stronger than the limits coming
from the Higgs searches.

\begin{figure}
\psfig{figure=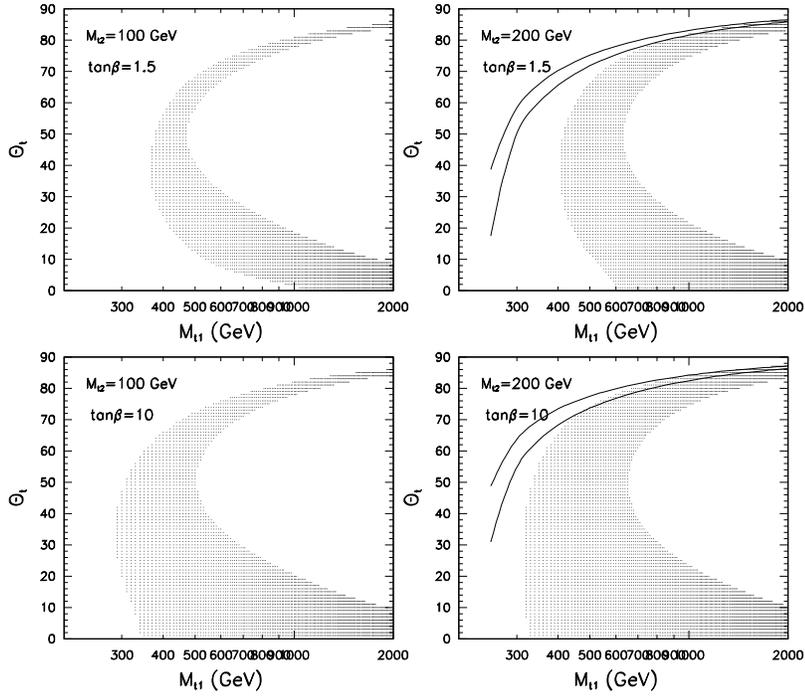,height=10.cm}
\caption{Allowed by $\Delta\chi^2<4$ and $M_h>70$ GeV regions in the plane 
$(M_{\tilde t_2},\theta_{\tilde t})$ for two different values of 
$M_{\tilde t_1}$ and two different values of $\tan\beta$. In the region 
between the two solid lines 170 GeV $<M_{\tilde b_1}<180$ GeV.} 
\label{fig:barc3}
\end{figure}

For fixed mass $M^2_{\tilde t_2}$ of one of the stops the precision data 
(and the direct Higgs boson mass limit) determine the allowed region in the 
($M^2_{\tilde t_1}$, $\theta_{\tilde t}$) plane shown in Fig. \ref{fig:barc3}
for two different values of $\tan\beta$ and two values of $M^2_{\tilde t_2}$.
The possibility of the existence of light, mostly left-handed stop (for
$\theta_{\tilde t}\simlt90^o$ the stop, whose mass is fixed on the Fig.
\ref{fig:barc3} to 100 or 200 GeV becomes almost purely left-handed), which
seems to contradict the standard argument that it should be heavy, can be
uderstood from the formula for their contribution to the $T$ parameter
\cite{HABER,MYDR} which (neglecting the mixing of sbottoms) reads:
\begin{eqnarray}
T = \cos^2\theta_{\tilde t}f\left(M^2_{\tilde t_1},M^2_{\tilde b_L}\right)
  + \sin^2\theta_{\tilde t}f\left(M^2_{\tilde t_2},M^2_{\tilde b_L}\right)
  - \sin^2\theta_{\tilde t}\cos^2\theta_{\tilde t} 
                           f\left(M^2_{\tilde t_1},M^2_{\tilde t_2}\right)
\nonumber
\end{eqnarray}
where $f(x,y)=
(3/16\pi s^2_WM^2_W)\left(x+y-{2xy\over x-y}\log{x\over y}\right)$ and 
$M_{\tilde b_L}$ is related to the physical stop masses through the relations
\begin{eqnarray}
M^2_{\tilde t_L}=\cos^2\theta_{\tilde t}M^2_{\tilde t_1}
                + \sin^2\theta_{\tilde t}M^2_{\tilde t_2}, ~~~
M^2_{\tilde t_R}=\sin^2\theta_{\tilde t}M^2_{\tilde t_1}
                + \cos^2\theta_{\tilde t}M^2_{\tilde t_2}\nonumber
\end{eqnarray}
and eq. (\ref{eqn:tbspl}). For $\theta_{\tilde t}\simlt90^o$
(but $\neq90^o$!) and $M_{\tilde t_1}$ large enough to make 
$M_{\tilde t_2}\approx M_{\tilde b_L}$ the negative third term in the 
expression for $T$ can cancel the first one and the second can also remain
small. From the above formulae it is clear, however, that those solutions
require $M_{\tilde t_R}\gg M_{\tilde t_L}$ which leads to very large 
inverse hierarchy $m^2_U\gg m^2_Q$ of the corresponding soft SUSY breaking
mass parameters. This is very unnatural from the point of view of the GUT 
boundary conditions which usually lead to $m^2_U\simlt m^2_Q$. It is also 
interesting to note that precisely this configuration is exploited 
in one of the most interesting attempts\cite{CADERAWA} to explain the 
anomalous HERA events\cite{HERA}. The mechanism proposed in\cite{CADERAWA}
requires mostly left-handed stop wih mass $\sim200$ GeV and (mostly 
left-handed) $\tilde b$ with $170<M_{\tilde b}<180$ GeV. From Fig. 
\ref{fig:barc3} where this $\tilde b$ mass range is marked by solid lines
one can judge how likely this (otherwise very appealing) solution to the
HERA puzzle is.

Another process which can further constrain the MSSM parameter space is the 
$b\rightarrow s\gamma$ decay. Due to the recent progress\cite{MISIAK} the 
theoretical prediction for $BR(b\rightarrow s\gamma)$ is now available with 
reduced error of $\approx15$\%. The existing measurement\cite{CLEO} of 
$BR(b\rightarrow X_s\gamma)$ implies therefore the important correlation of 
the charged (or, by the relation $M^2_{H^+}=M^2_A+M^2_W$, $CP$-odd) Higgs 
boson mass and masses and compositions of the lighter stop and chargino. To 
understand them, it is important to remember that the charged Higgs 
contribution to the $b\rightarrow s\gamma$ amplitude has always the same sign 
as the SM one whereas the chargino-stop contribution to this amplitude may 
have opposite sign. Since the actually measured value of $BR(b\rightarrow 
s\gamma)$ is close to the SM prediction, SUSY and charged Higgs contributions 
must either be small by themselves or cancel each other to a large extent. 

\begin{figure}[htbp]
\begin{center}
\begin{tabular}{p{0.48\linewidth}p{0.48\linewidth}}
\mbox{\epsfig{file=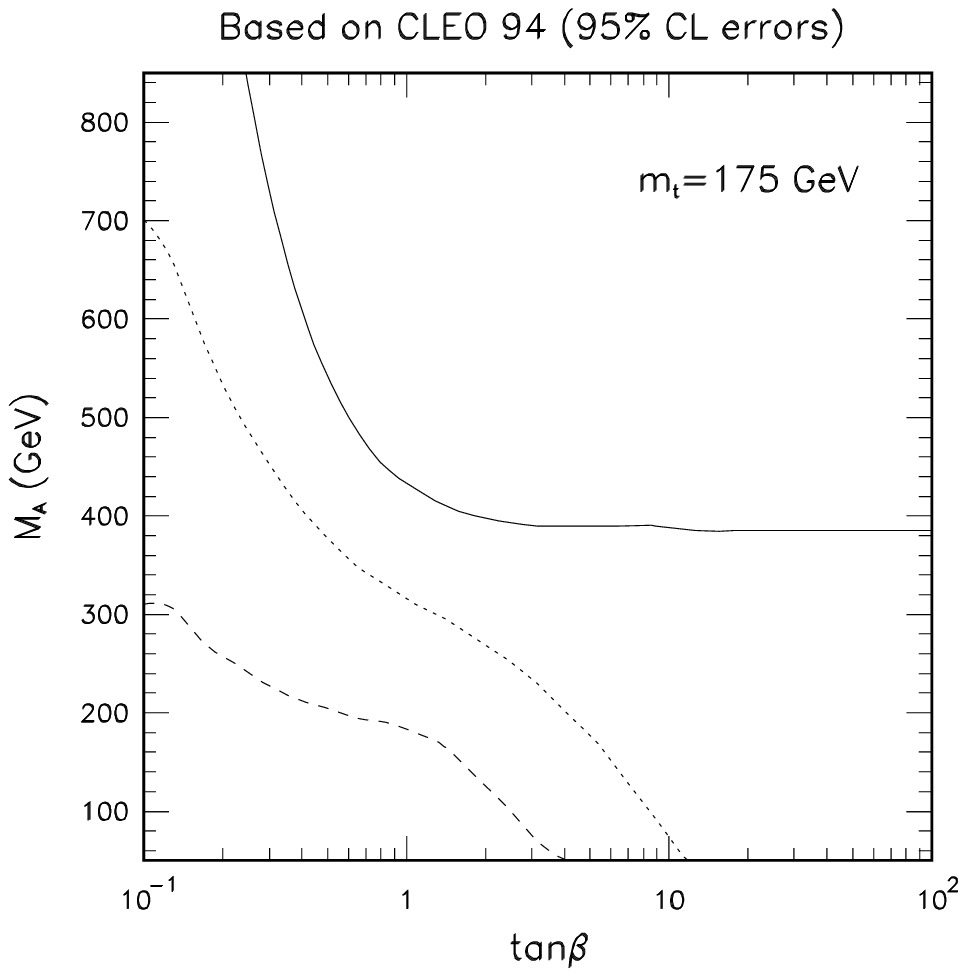,width=\linewidth}}
&
\mbox{\epsfig{file=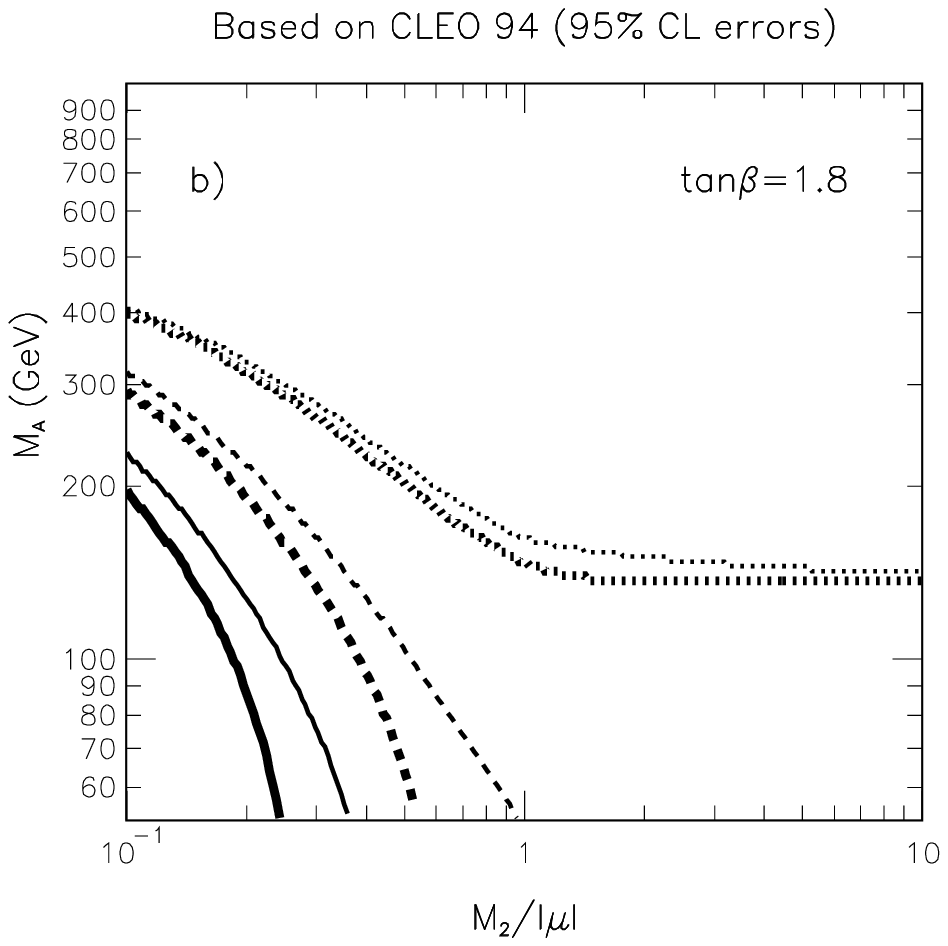,width=\linewidth}}
\\
\end{tabular}
\caption{{\bf a)} Lower limits on $M_A$ from $b\rightarrow s\gamma$ 
as a function of $\tan\beta$.
Solid line correspond to very heavy, $>{\cal O}(1$ TeV) sparticles.
Dashed (dotted) line show the limit for $m_{C_1}=M_{\tilde t_1}=$250 (500) 
GeV. {\bf b)} Lower limits on $M_A$ as a function of $M_2/|\mu|$,
based on CLEO $BR(B\rightarrow X_s\gamma)$ measurement. Thick lines
show limits for $\mu>0$, thin lines for $\mu<0$. Solid, dashed and
dotted lines show limits for lighter stop and chargino masses
$M_{\tilde t_1}=m_{C_1^\pm}=90$, 150 and 300 GeV,
respectively.\label{fig:barc45}}
\end{center}
\end{figure}

Fig. \ref{fig:barc45}a shows the lower limits on the mass of the 
$CP$-odd MSSM Higgs boson mass,
$M_A$, as a function of $\tan\beta$. Solid lines correspond to the 
case when all the superpartner masses are very large (above 1 TeV).  
Dashed (dotted) lines in Fig. \ref{fig:barc45}a show the same limits 
in the presence of chargino and stop with masses $m_{C_1}=M_{\tilde t_1}=500$
(250) GeV (with all other sparticles heavy) obtained by scanning over the 
values of $r=M_2/\mu$ and $\theta_{\tilde t}$. In the presence of light stop 
and chargino limits on $M_A$ are significantly weaker and totally disappear 
for large values of $\tan\beta$.
Another important observation is that, large chargino-stop contribution to 
$b\rightarrow s\gamma$ amplitude arise when the chargino is 
higgsino-like rather then gaugino-like i.e. when $M_2/|\mu | >1$. 
In addition, it depends also on the stop mixing angle $\theta_{\tilde t}$.  
Fig. \ref{fig:barc45}b (taken from ref. \cite{MIPORO}) shows the lower limit 
on the allowed pseudoscalar Higgs boson mass $M_A$ as a function of 
$r=M_2/|\mu|$ for three different values of the lighter chargino and lighter 
stop masses. For small $M_2/|\mu|$, i.e. 
for gaugino-like lighter chargino (when the chargino-stop contribution to 
$BR(b\rightarrow s\gamma)$ is suppressed) the resulting limits on $M_A$ are 
quite strong even for very light chargino and stop e.g.
$M_A\geq{\cal O}(200$ GeV) for $M_{\tilde t_1}=m_{C_1^\pm}=90$ GeV.
The limits decrease when $M_2/|\mu|$ increases and approximately saturate 
for $M_2/|\mu|\geq 1$. 

Other rare processes like e.g. neutral meson mixing ($K^0$-$\overline K^0$ 
or $B^0$-$\overline B^0$) etc. give much weaker constraints\cite{MIPORO}.
\vskip 0.3cm

\section{Constraints on specific supersymmetry breaking scenarios}
\vskip 0.2cm

Up to now we have been discussing bounds and correlations imposed by 
precision measurements on the parameter space of the unconstrained MSSM.
These bounds were ``absolute'' in the sense that e.g. to obtain lower limits
on slepton masses all other sparticles were kept heavy in order to reduce
their impact on the fit. In the following we shall impose the constraints
discussed in the preceding section on the specific models of supersymmetry
breaking, such as minimal SUGRA model or very popular recently gauge mediated 
models. In those models the resulting at the electroweak scale MSSM parameters 
are strongly correlated by specific boundary conditions assumed at the GUT
or intermediate scales (for example, sleptons are always lighter than third
generation squarks). For this reason, low energy precision measurements
constrain such models very efficiently. We will ilustrate those effects below
by discussing the predictions of these models for the mass of the lighter MSSM
Higgs boson as this issue is of direct interest for physics at LEP2. 
More detailed discussion can be found in \cite{KANE1}.
In this context there has also been often addressed the question of 
fine-tuning in the Higgs potential in models with the soft terms generated
at large scales \cite{DIGI}. Indeed, if supersymmetry is to be the solution to
the hierarchy problem in the SM, it should not introduce another fine-tuning
in the Higgs potential. ``Naturalness'' of a given parameter set can be 
quantified e.g. by calculating 
the derivatives of $M^2_{Z^0}$ with respect to the GUT (or intermediate) scale 
parameters of the model
\cite{DIGI}:
\begin{eqnarray}
\Delta_i\equiv\mid{a_i\over M_Z^2}{\partial M^2_Z\over\partial a_i}\mid
\label{eqn:der}
\end{eqnarray}
In the global analysis, which combines the electroweak breaking with 
experimental constraints, it is interesting to check the ``naturalness'' 
of different parameter regions i.e. to check the values $(\Delta_i)$
for each parameter set. Before presenting the results for the specific models,
one has to remember that the ``naturalness'' is only a qualitative 
criterion. Compared to the fine-tuning of many orders of magnitude in the SM,
cancellations of two or even three orders of magnitude are still very small. 
Secondly, the as yet unknown fundamental theory of SUSY breaking may predict
soft supersymmetry breaking terms correlated to each other, thus ``explaining''
the cancellations between them. We now present the results of our
global analysis, for the three scenarios considered and for several values
of $\tan\beta$. In each case the lightest Higgs boson mass is shown as a
function of the heavier stop.

\begin{figure}
\psfig{figure=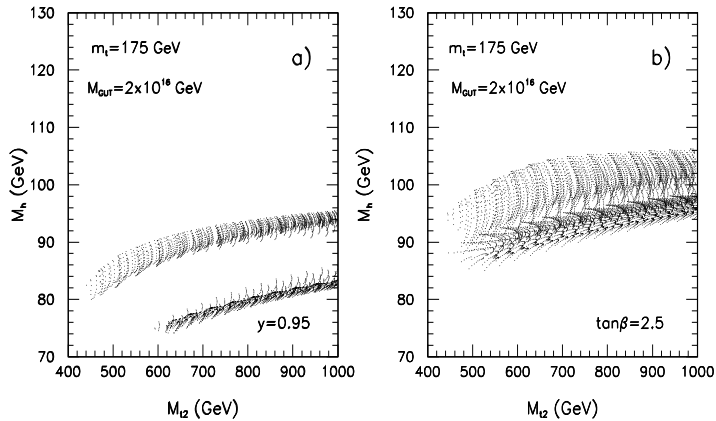,width=10.0cm,height=6.0cm}
\caption{$M_{h^0}$ versus the mass of the heavier stop ($M_{\tilde t_2}$)
obtained from universal boundary conditions at the scale $2\times10^{16}$ GeV
{\bf a)} close to the Q-IR fix point of the top quark mass ~
{\bf b)} for $\tan\beta=2.5$. In both cases
no fine-tunig criterion is imposed.
\label{fig:fintu1}}
\end{figure}

In the simplest, the so-called minimal supergravity model ({\sl Ansatz}), 
all superpartner masses are given in 
terms of five parameters: $m_0^2$, $M_{1/2}$, $\mu$, $A_0$ and $B_0$. Two of
them can be traded for $M_{Z^0}$ and $\tan\beta$ (for analytical solution of
the RG equations see \cite{CACHOLPOWA}). Thus, we get
strongly correlated superpartner spectrum and correlated with the Higgs boson
masses. It is now particularly simple to follow our global analysis and to
determine the allowed range of the lightest
Higgs boson mass as a function of the heavier stop mass.
In Fig. \ref{fig:fintu1} we show the results for $\tan\beta=1.65$ 
(corresponding to the infrared fixed point for the top quark Yukawa coupling)
and 2.5. We see that, in this 
model, requiring the proper breaking of the electroweak symmetry
and with the imposed experimental constraints the lightest Higgs 
boson mass is bounded from below: $M_{h^0}\simgt75(85)$ GeV for 
$\tan\beta=1.65(2.5)$ (for $\tan\beta=10$ the lower bound is around 105 GeV). 

The model also gives lower bound on $M_{A^0}$ of about 500 GeV at 
$\tan\beta=1.65$ and decreasing to 300 GeV at $\tan\beta=10$. The heavier stop 
is bounded from below at $\sim450$ GeV. Of course, the crucial role in 
obtaining those bounds is played by the universality {\sl Ansatz} combined 
with the existing experimental constraints. 
The mass $M_{h^0}$ is bounded from above at 95, 105 and 120  GeV for
$\tan\beta=1.65$, 2.5 and 10, respectively. Thus the general upper bound
on $M_{h^0}$ can be reached even in this constrained model.

\begin{figure}
\psfig{figure=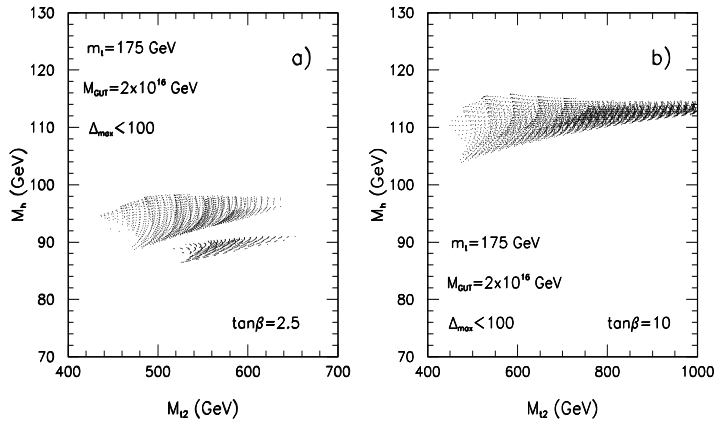,width=10.0cm,height=6.0cm}
\caption{$M_{h^0}$ versus the mass of the heavier stop ($M_{\tilde t_2}$)
obtained from universal boundary conditions at the scale $2\times10^{16}$ GeV
and requiring $\Delta_i<100$ 
{\bf a)} for $\tan\beta=2.5$ ($y\approx0.80$) ~
${\bf b)}$ for $\tan\beta=10$ ($y\approx0.71$).
\label{fig:fintu2}}
\end{figure}

Turning now to the fine-tuning problem we observe first that the model does
not admit at all solutions with all $\Delta_i<10$. This is mainly because 
of the imposed experimental limit $m_{C^\pm}>90$ GeV \cite{DIGI} which
pushes $M_{1/2}$ into the region with $\Delta_{M_{1/2}}\simgt10$ for all 
$\tan\beta$ values. Moreover, close to the IR fixed point (for 
$\tan\beta\approx1.65$), there do not even exist solutions with all
$\Delta_i<100$. As expected from the general arguments, cancellations become 
weaker with increasing $\tan\beta$. In Fig. \ref{fig:fintu2}a (b) we show the 
results for $\tan\beta=2.5(10)$ with the cut $\Delta_{max}<100$. We note that 
in this case such a cut leaves a non-empty parameter region but gives stronger
upper bounds on the Higgs boson mass for the same values of $M_{\tilde t_2}$. 
They result mainly from the bound on $A_0$ (i.e. on the left-right mixing)
obtained due to increasing $\Delta_{M_{1/2}}$ with increasing $A_0$.
Moreover, the cut $\Delta_{max}<100$ gives also an upper bound on 
$M_{\tilde t_2}$.  Finally we note one more interesting
effect: a cut on $\Delta$'s gives almost flat (instead of logarithmic)
dependence of $M_{h^0}$ on $M_{\tilde t_2}$. An increase in
$M_{\tilde t_2}$ is balanced by a decrease in $A_0$ (i.e. in the
left-right mixing) to keep $\Delta$'s below the imposed bound.

In the next step one can study a less restrictive model, with the pattern
of the soft terms consistent with the $SO(10)$ unification, i.e. with the
universal sfermion masses and the two Higgs boson masses taken as 
independent parameters. It turns out that the predictions the $M_{h^0}$,
$M_{A^0}$ and $M_{\tilde t_2}$ (as well as the degree of fine-tuning)
are very similar to those in the universal case and need not be independently
shown here. This can be partly understood in terms of the 
important role played by the limit $m_{C^\pm}>90$ GeV and by the 
constraints from $b\rightarrow s\gamma$ and precision data, which
are not sensitive to the assumed non-universality in the Higgs boson mass
parameters. Moreover, there is no change in the values of $\Delta_i$'s
since their expected decrease with the increasing number of free 
parameters is now reduced by the dissapearance of certain cancellations
in $\Delta_i$'s which are present in the universal case. Thus, 
the results presented in Figs. \ref{fig:fintu1} and \ref{fig:fintu2} are 
representative also for the partial breakdown of universality.

Finally, it is interesting to compare the supergravity scenario with models 
in which supersymmetry breaking is transmitted to the
observable sector through ordinary $SU(3)\times SU(2)\times U(1)$ 
gauge interactions of the so-called messenger fields at scales $M\ll M_{GUT}$
(see eg. \cite{ALEX}).
In general, these gauge-mediated models of SUSY breaking are characterized
by two scales: the scale $M$, which is of the order of the average messenger
mass and the scale $\sqrt{F}$ ($\sqrt{F}<M$) of supersymmetry breaking.
Messenger fields are assumed to form complete {\bf 5}+$\overline{\rm\bf 5}$ 
(or {\bf 10}+$\overline{{\rm\bf 10}}$)  $SU(5)$ representations. 
Their number $n$ is restricted to $n\leq4$ by  
perturbativity of the gauge couplings up to the GUT scale.

In terms of $M$ and $x\equiv F/M^2$ the soft supersymmetry breaking parameters 
of the MSSM at the scale $\sim M$ are given by:
\begin{eqnarray}
M_i = {\alpha_i(M)\over4\pi}M ~n ~x ~g(x) \equiv {\alpha_i(M)\over4\pi}M ~y
\end{eqnarray}
\begin{eqnarray}
m^2_{\tilde f} = 2M^2~n ~x^2 ~f(x) 
\sum_i \left({\alpha_i(M)\over4\pi}\right) C_i
=2M^2 ~y^2 ~z \sum_i \left({\alpha_i(M)\over4\pi}\right) C_i
\label{eqn:scalars}
\end{eqnarray}

\noindent
where $C_3=4/3$, $C_2=3/4$, $C_1=(3/5)Y^2$ ($Y$ being the hypercharge
of the scalar ${\tilde f}$), the functions $g(x)$ and
$f(x)$ ($g(0)=f(0)=1$, $g(1)\approx 1.4$, $f(1)\approx0.70$) 
can be found in ref. \cite{DIGIPO} and the factor $z\equiv f(x)/n g^2(x)$. 
Thus, for fixed messenger sector
(i.e. fixed $n$) and fixed scale $M$ all soft supersymmetry breaking masses
are predicted in terms of $y$ ($0<y<n_{max}~g(1)\approx5.6$) 
In those models we also have $A_0\approx0$ as the $A_0$ parameter can be
generated at two loop only \cite{DITHWE}. 
However, the values of the soft masses $m^2_{H_{1,2}}$ may differ 
significantly from their values given by eq. (\ref{eqn:scalars}) since they
can be modified by physics involved in generation of $B_0$ and $\mu_0$
parameters \cite{DVGIPO}. Therefore, in our scans we take $y$, $m_{H_1}$,
$m_{H_2}$, $\mu_0$ and $B_0$ as free parameters (the last two are fixed
by $M^2_{Z^0}$ and $\tan\beta$). To be general, the factor $z$ in eq. 
(\ref{eqn:scalars}) is scanned between $z_{min}=f(1)/n_{max}g^2(1)\approx0.15$ 
and $z_{max}=1$. For definitness we will consider $M=10^5$ GeV only.

We follow the same simple approach we used for the supergravity models. 
On the parameter space consistent with the electroweak symmetry breaking we
impose the discussed earlier experimental constraints
(now we require $m_{C^\pm}>120$ GeV, $M_{\tilde t_1}>140$ GeV). 
Very important r\^ole is played by $b\rightarrow s\gamma$. The requirement 
of good $b\rightarrow s\gamma$ rate reduces otherwise rather 
widely spread out $h^0$ and $A^0$ Higgs boson masses (for $\tan\beta=2.5$: 
$20<M_{h^0}<100$ GeV) to a narrow band ($80<M_{h^0}<100$ GeV, $M_{A^0}>200$ 
GeV). This effect can be easily understood (see Fig. \ref{fig:barc45}a) 
because in the model considered squarks and charginos are rather heavy 
\footnote{In addition, because $\mu$ values required by electroweak
symmetry breaking are large, the lighter chargino turns out to have only
small higgsino component and hence its $b \tilde t C^-$ coupling
is weaker than that of the pure higgsino chargino which is responsible for  
the limit shown in Fig. \ref{fig:barc45}a.} so a light $A^0$ is not allowed
by $b\rightarrow s\gamma$
and light $h^0$ is always associated with light $A^0$. Moreover, surviving 
small values of $M_{h^0}$ ($\sim80$ GeV for $\tan\beta=2.5$ are associated 
with lowest values of $M_{\tilde t_2}$ ($\simlt 500$ GeV) which are eliminated 
by imposing the $\Delta\chi^2<4$ cut. 
Finally, if we also require ``naturalness'' e.g. by demanding 
$\Delta_{max}<100$  we constrain the heavier stop mass 
$M_{\tilde t_2}$ and $CP$-odd higgs boson mass $M_{A^0}$ from above 
to $\simlt700$ GeV. 

\begin{figure}
\psfig{figure=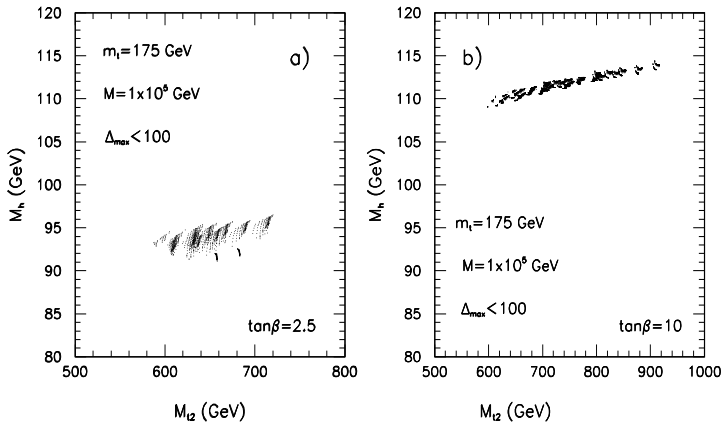,width=10.0cm,height=6.0cm}
\caption{Results for gauge mediated supersymmetry breaking models with 
$M=10^5$ GeV for $\tan\beta=2.5$: ~a) 
$M_{h^0}$ versus the mass of the heavier stop ($M_{\tilde t_2}$)
${\bf b)}$ versus the mass of the $CP$-odd Higgs boson ($M_{A^0}$).
Condition $\Delta_i<100$ is imposed. 
\label{fig:fintu3}}
\end{figure}

Final results are shown in 
Fig. \ref{fig:fintu3} as a plot of $M_{h^0}$ 
versus the mass of the heavier stop $M_{\tilde t_2}$ predicted in models 
of gauge mediated supersymmetry breaking with $M=10^5$ GeV for 
$\tan\beta=2.5$ and 10. 
As in the case of supergravity models, the restriction of the chargino and 
stop masses eliminates solutions with $\Delta_{max}<10$. 
With all constraints imposed, 
$M_{h^0}$ turns out to be surprisingly tightly constrained. 
For $\tan\beta=2.5(10)$ values of the lightest scalar Higgs boson
are bounded by 90(108) GeV $\simlt M_{h^0}\simlt97(115)$ GeV.
Masses of the $CP$-odd Higgs boson are bounded to 280(200) GeV
$\simlt M_{A^0}\simlt$700(850) GeV. These upper bounds should be compared 
to the ones obtained in \cite{RITOMO} in the restricted model of
gauge mediated supersymmetry breaking (with $x=1$, $n=1$ and with
$m_{H_{1,2}}$ as given by eq. (\ref{eqn:scalars})) without imposing any
additional constraints. It is interesting that in the much more general 
scenario described above, after imposing experimental and 
naturalness cuts, 
one gets upper bounds on $M_{h^0}$ not higher than those obtained 
in \cite{RITOMO}.

\section{Summary}

There is an apparent contradiction between the hierarchy problem (which 
suggest new physics to be close to the electroweak scale) and the striking 
success of the Standard Model in describing the electroweak data. The 
supersymmetric extension of the SM offers an interesting solution to this 
puzzle. The bulk of the electroweak data is well screened from supersymmetric 
loop effects, due to the structure of the theory, even with superpartners 
light, ${\cal O}(M_Z)$. In order to maintain the overall succes 
of the SM, only left-handed squarks from the third generation have to be 
$\simgt {\cal O}(300$ GeV). Some, rather weak limits do exist also for 
left-handed sleptons. The other superpartners can still be light,
at their present experimental mass limits, and would manifest themselves 
through virtual corrections to the small number of observables such as 
$b\rightarrow s\gamma$, $K^0$-$\bar K^0$ and  $B^0$-$\bar B^0$ mixing and a 
few more for large $\tan\beta$ (not discussed in this talk)\cite{SOLA,CAGIWA}.

In unconstrained minimal supersymmetric extension of the Standard Model
there exists the well known upper bound on the mass of the lightest 
supersymmetric Higgs. The available parameter space can be efficiently 
reduced by additional theoretical assumptions, related to the 
extrapolation of the model to very large energy scales. When these 
assumptions are supplemented by the low energy experimental constraints
and qualitative criterion of ``naturalness'' more more definite predictions
for $M_{h^0}$ are obtained. In this context, it is interesting to note that
in the models considered $M_{h^0}>75$ GeV (roughly the current experimental 
direct bound). Moreover, several arguments point toward $M_{h^0}<$100 GeV. 
Both, the discovery or the absence of the Higgs boson in this mass range will 
have strong implications for the supersymmetric extension of the Standard
Model.
\vskip 0.2cm

{\bf Acknowledgments}
This work was supported by Polish State Commitee for Scientific Research under
grant 2 P03B 040 12 (for 1997-98).

\end{document}